\documentclass[manuscript, nonacm]{acmart}

\usepackage{subcaption} 
\usepackage{microtype}
\usepackage{listings}
\usepackage{multirow}
\usepackage{makecell}
\usepackage{balance}
\usepackage{rotating}
\usepackage{url}
\usepackage{setspace}
\usepackage{fancyvrb}
\usepackage{soul}
\usepackage[T1]{fontenc}
\usepackage{lmodern}

\begin{document}
\title{A R4RS Compliant REPL in 7 KB}



\author{Léonard Oest O'Leary}
\affiliation{
	\institution{Universit\'e de Montr\'eal}            
	\country{Canada}                    
}
\email{leonard.oest.oleary@umontreal.ca}          

\author{Mathis Laroche}
\affiliation{
	\institution{Universit\'e de Montr\'eal}            
	\country{Canada}                   
}
\email{mathis.laroche@umontreal.ca}         

\author{Marc Feeley}
\affiliation{
	\institution{Universit\'e de Montr\'eal}            
	\country{Canada}                   
}
\email{feeley@iro.umontreal.ca}         


\begin{abstract}

  The Ribbit system is a compact Scheme implementation running
  on the Ribbit Virtual Machine (RVM) that has been ported to a dozen host languages. 
  It supports a simple Foreign Function Interface (FFI) allowing extensions 
  to the RVM directly from the program's source code. We have extended the system 
  to offer conformance to the R4RS standard while staying as compact as possible. 
  This leads to a R4RS compliant REPL that fits in an 7 KB Linux executable. This 
  paper explains the various issues encountered and our solutions to make, arguably,
  the smallest R4RS conformant Scheme implementation of all time. 

\end{abstract}



\keywords{Virtual Machines, Compiler, Dynamic Languages, Scheme, Compactness}  

\maketitle

\section{Introduction}\label{sec:intro}

The Ribbit Scheme system~\cite{Ribbit1,Ribbit2} is portable, extensible, and compact. It is
based on a Virtual Machine (VM) that is portable to a dozen host languages including: 
JavaScript, C, Assembly (x86), Shell, Haskell, and Prolog. It is extensible, enabling
programmers to add their own host-level primitives in Scheme code or using annotations
within the VM's code. It is compact by design, with an extremely simple VM and with an AOT compiler that
removes dead code from the program, library, and VM itself. 


This paper explains how Ribbit has been extended to maintain a small size while
adding conformance to the {R4RS} specification. The main enhancements to the previous Ribbit system are:
\begin{enumerate}
\item Support for variadic procedures and rest parameters.
\item Implementation of all required file I/O procedures.
\item Various measures to better compact the generated code,
including a new approach for encoding programs and a compact implementation of the standard library.
\end{enumerate}
These changes have allowed us to fit an interactive REPL fully conforming to R4RS in a 7 KB Linux executable program with no external dependencies. We chose to support the R4RS Scheme standard because
it combines practicality and small size. Also, there is lots of existing code that can run in an
R4RS system including most of the SLIB Portable Scheme Library~\cite{SLIB}. Subsequent standards added
features that would increase the size of the system substantially:
hygienic macros and multiple values are required starting at R5RS, and libraries and Unicode
support are required starting at R6RS. A more detailed reasoning for our choice
can be found in Section~\ref{sec:r4rs}. 

The paper is organized as follows:
Section~\ref{sec:ribbit} provides an overview of the Ribbit system.
Section~\ref{sec:encoding} explains the encoding optimizations. Section~\ref{sec:r4rs} describes the
implementation of the {R4RS} library to achieve compactness and portability across host languages.
Section~\ref{sec:x86-host} describes the x86 assembly host which is our most compact and fast implementation of the RVM.
Section~\ref{sec:evaluation} evaluates the effectiveness of our approach through benchmarks that measure the space and execution time using multiple compilation settings.
Finally, the paper concludes with related work.

\section{Ribbit}\label{sec:ribbit}

Ribbit has three main components: the Ribbit VM (RVM) implemented in multiple host languages,
the Ribbit Scheme Compiler (RSC), and the standard library. RSC, an Ahead Of Time (AOT)
compiler, combines the source program with the standard library to generate a standalone
specialized RVM in the chosen host language. Every RVM source program contains annotations that attach
meaning to portions of its code. This lets the compiler selectively
include, exclude or adapt sections of the code leading to a RVM uniquely tailored
to the program.

The compiler will embed in the RVM source code the RVM code it has generated for the
program in an encoded form:
the Ribbit Intermediate Byte Notation (RIBN), pronounced \textit{ribbon}. The RIBN has
two parts: the symbol table and the encoded sequence of RVM instructions. The symbol table is represented as a list, where 
the position of a symbol in the list is its index. When encoding the symbol table inside the RIBN, the list, as well as the string representation of symbols, are encoded 
in reverse order for decoding simplicity. The encoded program uses a specialized encoding discussed
in Section~\ref{sec:encoding}.

\subsection{Ribbit VM}

The Ribbit VM was designed with simplicity in mind, to minimize the VM's code size and allow
porting it to new host languages with low effort. It is a stack machine with 6 available instructions
loosely corresponding to the fundamental Scheme constructs:
\texttt{jump} (tail call), \texttt{call} (non-tail call), \texttt{set} (writing a variable),
\texttt{get} (reading a variable), \texttt{const} (literal data), and \texttt{if} (conditional
execution).

To simplify memory management, the only data that is managed by the RVM is the \textit{rib}:
a three field structure where each field can be an integer or a reference to a rib.
The code executed by the RVM, the Scheme data, and the stack are all represented using ribs.
When a rib represents Scheme data, the last field is an integer indicating the type:
0 for pair, 1 for procedure, 2 for symbol, 3 for string, etc. In the rest of the paper we
will use the notation \texttt{[\textit{a},\textit{b},\textit{c}]} to mean a rib with the fields
\texttt{\textit{a}}, \texttt{\textit{b}}, and \texttt{\textit{c}}. This also happens to be the implementation of ribs in the
Python and JavaScript RVMs, among others. As an example, the Scheme improper list \texttt{(1~2~.~3)}
is represented using two ribs: \texttt{[1, [2, 3, 0], 0]}.

\begin{figure}
\begin{center}
\begin{minipage}{3in}
\begin{verbatim}
(lambda (n)  ;; hypothetical definition of abs
  (let ((sign (if (< n 0) -1 1)))
    (* sign n)))
\end{verbatim}
\end{minipage}
\end{center}
\vspace{1em}
    \includegraphics[scale=0.27]{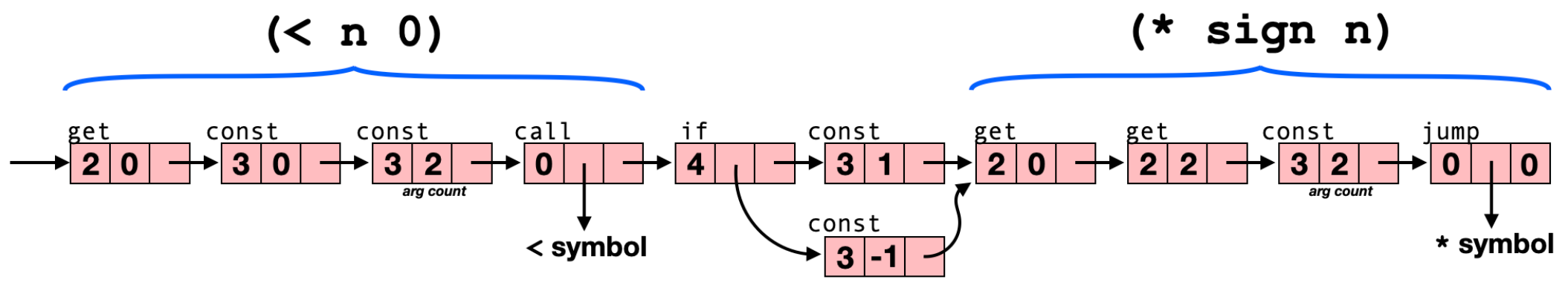}
    \caption{RVM code represented as ribs. The RVM code corresponds to the body of the lambda-expression.}
    \label{fig:abs}
\end{figure}

Global variables are implemented
by storing the variable's value in the first field of the symbol naming the variable. 
The second field of a symbol contains the string representation of this symbol. If the symbol
is anonymous, meaning that its string representation is not needed, this field is empty.
The RVM code is stored in memory as a chain of ribs linked using the third field. The first
field is the opcode, an integer indicating the instruction type.
The second field is the operand. For 
\texttt{jump}, \texttt{call}, \texttt{set}, and \texttt{get} instructions it indicates the location
of a cell (either using a symbol if it refers to a global variable, or an integer if it
is a stack slot). For the \texttt{const} instruction the operand is the literal value.
For the \texttt{if} instruction the operand is the next instruction to execute if the value
popped from the stack is not \texttt{\#f}. Figure~\ref{fig:abs} gives an example of Scheme code
and the equivalent RVM code representation as ribs.
Note that both \texttt{jump} and \texttt{call} have the same opcode (0). They are distinguished
by the third field which is 0 in the case of a \texttt{jump} (i.e.~there is no following
instruction).

When the RVM is executed it decodes the RIBN to create the symbol table and
the RVM code represented as ribs, and then executes that code. The RIBN decoding
is discussed in further detail in Section~\ref{sec:encoding}.

\subsection{Ribbit Scheme Compiler}

Ribbit's AOT compiler merges the standard library and the source program to perform
a whole-program analysis. A liveness analysis removes unused procedures and
primitives to optimize compactness. Annotations let the compiler remove,
reorder and add new primitives. If the liveness analysis detects that a certain
primitive is not used, then it is removed from the RVM source code.

\subsection{Annotations}\label{sec:annotation}

Annotations live inside the RVM's source code and give additional information
to the compiler to generate a specialized VM. They have a syntax similar to s-expressions,
but start and end with \texttt{@@(} and \texttt{)@@} to easily embed them
unambiguously in host language comments. Annotations are composed of a name
and $\geq 0$ arguments and refer to some section of the host code. If the
annotation starts and ends on the same line the annotation refers to the
code on that line. For example, in the C RVM, there is the following inline \texttt{feature}
annotation that refers to the \texttt{\#include} line: 

\begin{center}
\begin{figure*}[ht]
\begin{Verbatim}[frame=single]
#include <stdio.h> // @@(feature stdio)@@
\end{Verbatim}

\caption{Example of a \texttt{feature} annotation inside the C RVM}
\label{ex:rvm_c_annotations_feature}
\end{figure*}
\end{center}

If the annotation spans multiple lines, then the annotation refers to the code
from the start line to the end line inclusively. Annotations can also be nested. For
instance, the \texttt{primitives} (plural) annotation includes multiple
\texttt{primitive} (singular) annotations. This lets the compiler know the set of
primitives implemented by the RVM and where each primitive is implemented. As
an example, here is a multiline \texttt{primitive} nested inside a
\texttt{primitives} annotation, as is found in the primitive procedure
dispatch logic of the C RVM:

\begin{figure*}[ht]
	
\begin{Verbatim}[frame=single]
switch (prim_index) {
// @@(primitives (gen "case " index ":" body)
...
    case 19: // @@(primitive (putchar c) (use stdio)
      putchar(NUM(tos())); break; // print top of stack
    // )@@
...
// )@@
}
\end{Verbatim}
\caption{Example showcasing \texttt{primitive} and \texttt{primitives} annotations inside the C RVM}
\label{ex:rvm_c_annotations_primitives}
\end{figure*}

\texttt{feature} annotations are used to control the inclusion of various
parts of the RVM, some of which may not be needed for a given source program.
The dependency of a feature on another feature is indicated by the \texttt{use}
clause, as in the above \texttt{putchar} primitive that depends on the
\texttt{stdio} feature.
This lets the RSC compiler remove, add, and renumber the primitives inside the
RVM. In the previous example, the \texttt{putchar} primitive may get a
different index if other primitives are not needed or it may itself be removed
from the specialized RVM. 

The location where the RIBN needs to be injected into the RVM is indicated with the \texttt{replace} annotation. For example, the following code tells
the compiler to replace \texttt{"encoded RVM code"} by the result of
\texttt{(encode~92)}, the RVM code encoded as a string which is the plain
base 92 RIBN encoding:

\begin{figure*}[ht]
\begin{Verbatim}[frame=single]
// @@(replace "\"encoded RVM code\"" (encode 92)
ribn = "encoded RVM code"
// )@@
\end{Verbatim}
\caption{Example of a \texttt{replace} annotation inside the Python RVM}
\label{ex:rvm_c_annotations_replace}
\end{figure*}

\subsection{Features}

Features in Ribbit are compile time variables that enable the compiler
to fine-tune the RVM. They are defined in the RVM source code using the
\texttt{feature} annotation or in the Scheme source code using the
\texttt{define-feature} form. Note that primitives are also features, meaning
that when a primitive is enabled, the feature with the same name is enabled as
well and vice-versa. 
Features and primitives can be enabled or disabled in
multiple ways:

\begin{description}
\item[By the programmer using RSC command line options.] This is done with the command line options \\
    \texttt{-f+~feature-to-enable} or \texttt{-f-~feature-to-disable}. This allows
    fine tuning of the RVM, for example choosing whether the JavaScript RVM is to be
    run on the web or on NodeJS.

\item[With dependencies among the features.] Features can define dependencies
    with other features with the \texttt{(use~...)} clause. A fix-point
    algorithm is used to determine the set of features to enable or disable.

\item[By the compiler.] If the compiler detects certain optimizations, it can enable or disable
    features. For example, the \texttt{arity-check} and \texttt{rest-param} features are 
    enabled if the compiler detects that the source program has \texttt{lambda} expressions
    with rest parameters (after dead code elimination).

\end{description}

Ribbit's extensibility is mainly achieved using
a simple Foreign Function Interface (FFI) to the host language
through the \texttt{define-primitive} form.
It defines a primitive with a name, a body and
an optional \texttt{use} clause. The body is a string containing host-language
code and is injected inside the RVM source code. The \texttt{use} clause indicates
dependencies between features. For example, the \texttt{putchar} primitive
depends on the \texttt{stdio} feature as it needs C's \texttt{putchar} function. 
The use of the \texttt{define-primitive} form below is equivalent to the annotations within the RVM shown in Figure~\ref{ex:rvm_c_annotations_primitives}.

\begin{figure*}[ht]
\begin{Verbatim}[frame=single]
(define-primitive (putchar c)
  (use stdio)
  "putchar(NUM(tos())); break;")
\end{Verbatim}
\caption{Example showcasing the definition of a primitive in Scheme code.}
\label{ex:macro_primitive}
\end{figure*}

In a similar way, a \texttt{define-feature} form exists and lets
programmers add functionality to the RVM through code embedded inside the RVM.
This embedded code can be enabled or disabled depending on the state of the feature.
When using \texttt{define-feature}, one needs to specify the location where the code must
go. These locations are identified using \texttt{@@(location} \emph{name}\texttt{)@@} annotation. 
For instance,  the following feature definition, created using the \texttt{define-feature} form, is equivalent to the one in Figure~\ref{ex:rvm_c_annotations_feature} . Here \texttt{decl} is a named location present at the beginning of the C RVM file.

\begin{figure*}[ht]
\begin{Verbatim}[frame=single]
(define-feature (stdio)
  (decl "#include <stdio.h>"))
\end{Verbatim}
\caption{Example showcasing the definition of a feature in Scheme code.}
\label{ex:macro_feature}
\end{figure*}

\label{sec:ribbi_extension_replace}
\subsection{Extension of the Replace Annotation}

As will be discussed in Section~ \ref{sec:encoding}, Ribbit now supports a 
base 256 RIBN encoding. 
For compiled hosts such as C and x86 assembly, encoding the RIBN as a constant array of bytes is
the most compact approach. The \texttt{replace} annotation has been extended to support the embedding
of a literal array through the \texttt{(encode-as-bytes} \emph{RIBN-base} \emph{prefix} \emph{separator} \emph{suffix}\texttt{)} procedure. 
For example, the C RVM embeds the base 256 RIBN in this way:
\\

\begin{Verbatim}[frame=single]
// @@(replace "literal-array-to-embed" (encode-as-bytes 256 "{" "," "}")
unsigned char compressed_ribn[] = literal-array-to-embed;
// )@@
\end{Verbatim}

\noindent
The corresponding line of the generated RVM will look like this:

\begin{Verbatim}[frame=single]
unsigned char compressed_ribn[] = { 41, 59, 39, 117, 63, 62, ... };
\end{Verbatim}

More broadly, the \texttt{replace} annotation has been extended to embed information known at compile time and needed
by the RVM at run time. For example, this is used during the compression (see Section~\ref{sec:encoding_lzss}) and the
specialized encoding (see Section~\ref{sec:encoding_specialization}). To embed such information inside the RVM,
\texttt{features} can now contain values, such as lists, numbers, and other Scheme values. This information is
accessible through the \texttt{replace} annotation. Combined with the use of procedures that convert the feature-values
into strings, this information can be embedded inside RVMs easily. For example, this mechanism is used by the C RVM
to create an uninitialized array of the exact size of the decompressed RIBN:

\begin{Verbatim}[frame=single]
// @@(replace "RIBN_SIZE" compression/lzss/2b/ribn-size
unsigned char ribn[RIBN_SIZE]; 
// )@@
\end{Verbatim}

\noindent
Here, the feature \texttt{compression/lzss/2b/ribn-size} contains the uncompressed size of the RIBN. Without an
information sharing mechanism between the compiler and the RVM, one would have to resort to dynamically allocated vectors, complicating the logic and increasing the footprint of the RVM.

\subsection{\texttt{if-feature} Form}

There are instances where Scheme code must behave differently based on features being enabled or disabled. 
This is the case for the \texttt{eval} procedure of the standard library that depends on the \texttt{arity-check} feature. The \texttt{arity-check} feature tells the compiler to add support for argument count verification: pushing the number of arguments onto the stack just before a \texttt{call} or \texttt{jump} (which is otherwise not needed). This enables the RVM to check that the number of arguments matches the arity of the procedure. In other words, the \texttt{arity-check} feature impacts the calling protocol chosen by the compiler and \texttt{eval} needs to be aware of it.

The special form \texttt{if-feature} was added to test the use of specific features. This special form is processed after the \textit{liveness} analysis. This timing is essential because determining the liveness of a feature is dependent on the \texttt{define-primitive} and \texttt{define-feature} forms, which use \texttt{use} clauses to indicate dependencies.

\section{Encoding}\label{sec:encoding}

The explicit chaining in the RVM code's rib representation allows
representing loops (cycles) and join points (sharing)
without additional instructions. Although the compiler does not
take advantage of this for loops, it does use sharing for
the join points of non-tail \texttt{if} forms, as in
Figure~\ref{fig:abs}. So the rib representation of the RVM code is a
Fork-Join Directed Acyclic Graph (DAG) with optional joins that we will call the \textit{code graph}.

The RIBN is an encoding of the code graph generated by the compiler
that is decoded by the RVM to create the code graph that the RVM
interpreter uses. The RIBN is conceptually a list of integer codes
whose values are in the range 0 to $rb-1$, where $rb$ is the
\textit{RIBN base}.  The goal is to encode the code graph such that
the least space is taken by the sum of the RIBN and the implementation
of the decoder that is part of the RVM.  In the previous Ribbit system
the chosen encoding was suboptimal:

\begin{figure}
    \includegraphics[scale=0.27]{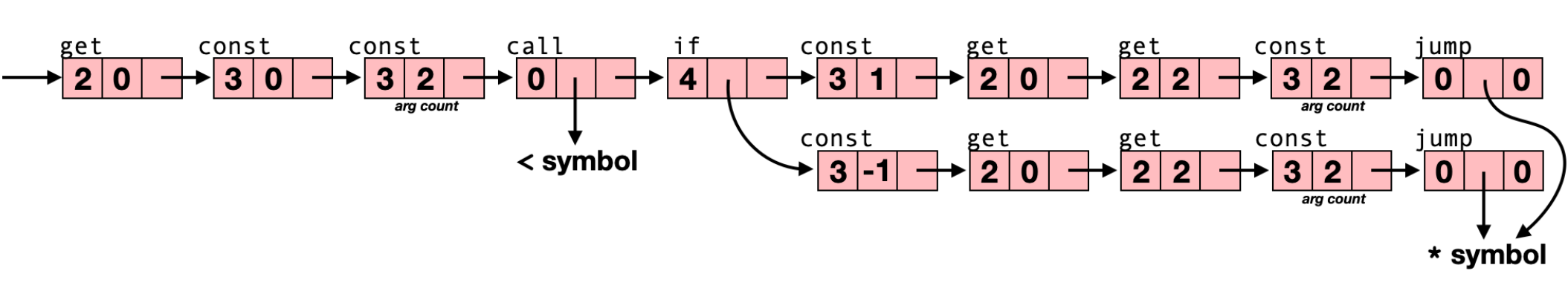}
    \caption{Code graph of Figure~\ref{fig:abs} after tail duplication transformation.}
    \label{fig:abs-tail-dup}
\end{figure}

\begin{enumerate}
\item
  The RIBN was a string of characters with a RIBN base equal to 92,
  the set of characters that don't require escaping in most host
  languages. In host languages that are compiled, where the system's footprint
  is the size of the executable, it is more space efficient to use an
  array of bytes and a RIBN base of 256. We solve this by allowing each
  host language to define the RIBN base and how the RIBN is stored in
  memory.  The compiler will transform the generated RIBN to comply with
  the defined encoding and map the RIBN codes appropriately, for example,
  between the 92 codes and the printable ASCII characters.
\item
  The RIBN could only express a tree structure. A tail duplication
  transformation was applied to the code to remove any shared
  structure.  For example, the code graph in Figure~\ref{fig:abs} was
  encoded in a RIBN that was decoded into the code graph in
  Figure~\ref{fig:abs-tail-dup}.  This duplication causes an
  exponential growth of the code when multiple non-tail \texttt{if}
  forms are in sequence. This is solved by the new encoding approach
  which can express sharing.
\item
  There was only one RIBN encoding supported. Now, different 
  encoding mechanisms exist. The \verb|optimal| encoding empowers 
  the compiler to decide on the optimal ranges to assign each decoding instruction.
   The corresponding
  decoder is used by the RVM generated by the compiler.
\item 
  No compression was applied to the RIBN before embedding it into the RVM. 
  Now, LZSS compression is applied on compatible hosts 
\end{enumerate}

\newcommand{\esc}[1]{\scalebox{1}{#1}}

\subsection{Decoding Instructions}\label{sec:runtimeinstructions}








The decoding of the RIBN into the code graph is done using a
stack. The RIBN is a sequence of \emph{decoding instructions} executed
by the decoder that have an effect on the stack. The stack is
initially empty and at the end of the decoding it contains a single
value that is the code graph.  The decoding instructions should not be
confused with the \emph{RVM instructions} that are contained in the
code graph and executed by the RVM's interpreter.

Because of the linked nature of the code graph, it is easier to construct
the code graph starting with the tail, for example starting with the
\texttt{jump} instruction of Figure~\ref{fig:abs} and then adding
the previous instructions in front one by one. Note also that all
code sequences must end in a \texttt{jump}, so a RIBN always starts
off with a decoding instruction for a \texttt{jump}.

Decoding instructions have a type and an argument (with one exception
for decoding \texttt{if}).  The argument is always a nonnegative
integer that either stands for itself (the \texttt{int} case), or is
used to index the symbol table to refer to a specific symbol (the
\texttt{sym} case).  To avoid arbitrarily limiting the number of
symbols and global variables of the source program, the argument has
no upper limit. Consequently each decoding instruction is a sequence
of one or more RIBN codes, using a variant of Variable-Length
Quantity \cite{VLQ} (VLQ).  VLQ uses fewer RIBN codes for the smaller
argument values. The variant we use combines in a single RIBN code the
decoding instruction type and the argument value when it is
small. When a decoding instruction is a single RIBN code it is called
the \emph{short} form, otherwise it is the \emph{long} form.  To
enhance compactness, the RSC compiler analyzes the program to assign
small symbol table indexes for the most frequently referenced global
variables, so as to minimize the frequency and length of the
\emph{long} form.

Ideally the set of decoding instructions would allow handling any code
graph. However, it simplifies the decoder to restrict the code graph
and this may reduce the size of the decoder in the generated RVM.

Firstly, the operand of \texttt{const} instructions is restricted to
be a symbol, a nonnegative integer, or a \emph{constant procedure}
(meaning with no free variables, which is useful for the frequent case
of top-level procedure definitions). The RSC compiler ensures that
this is the case by doing a rewriting of the code graph before the
RIBN is created. Any \texttt{const} instruction with an operand that
is not an acceptable constant is turned into a \texttt{get}
instruction that refers to a freshly created global variable that
contains the constant. The compiler also adds to the beginning of the
program the RVM instructions that constructs the constant and store it
in the global variable. This is done in a way that shares common parts
of constants, in particular when the same constant appears in several
places in the source code a single global variable is used.

\subsection{$SHARE$ Decoding Instruction}\label{sec:share}

The previous Ribbit had a second restriction on the code graph, namely
that it had to be a tree. Ribbit now supports DAGs thanks to a
decoding instruction called $SHARE$ that was not available
previously.

\begingroup
\newcommand{\htablesize}[0]{\vspace{5pt}}

\begin{table}[]

\begin{tabular}{llllrcl}
\begin{tabular}[c]{@{}c@{}}Decoding\\ instruction\\ type\end{tabular} & 
\begin{tabular}[c]{@{}c@{}}Argument\\ (\textit{arg})\end{tabular} &
\begin{tabular}[c]{@{}c@{}}RVM\\ instruction\\ generated\end{tabular} & &
\multicolumn{3}{l}{
\begin{tabular}[c]{@{}c@{}}Effect on decoding stack state using the notation\\$\langle$\textit{current-stack-state}$\rangle$ $\rightarrow$ $\langle$\textit{next-stack-state}$\rangle$\end{tabular}} \\
\htablesize
\\
\htablesize
$PUSH_0$   & \texttt{int} or \texttt{sym}  & \texttt{jump}    && \textit{stack}\ldots & $\rightarrow$ & \texttt{[0,} \textit{arg}\texttt{, 0]} \textit{stack}\ldots \\
\htablesize
$LINK_0$   & \texttt{int} or \texttt{sym}  & \texttt{call}    && \textit{x} \textit{stack}\ldots & $\rightarrow$ & \texttt{[0,} \textit{arg}\texttt{, }\textit{x}\texttt{]} \textit{stack}\ldots \\
\htablesize
$LINK_1$   & \texttt{int} or \texttt{sym}  & \texttt{set}     && \textit{x} \textit{stack}\ldots & $\rightarrow$ & \texttt{[1,} \textit{arg}\texttt{, }\textit{x}\texttt{]} \textit{stack}\ldots \\
\htablesize
$LINK_2$   & \texttt{int} or \texttt{sym}  & \texttt{get}     && \textit{x} \textit{stack}\ldots & $\rightarrow$ & \texttt{[2,} \textit{arg}\texttt{, }\textit{x}\texttt{]} \textit{stack}\ldots \\
\htablesize
$LINK_3$   & \texttt{int} or \texttt{sym}  & \texttt{const}   && \textit{x} \textit{stack}\ldots & $\rightarrow$ & \texttt{[3,} \textit{arg}\texttt{, }\textit{x}\texttt{]} \textit{stack}\ldots \\
\htablesize
$MERGE_3$  & \texttt{int}                  & \texttt{const}   && \textit{y} \textit{x} \textit{stack}\ldots & $\rightarrow$ & \texttt{[3, [[}\textit{arg}\texttt{, 0, }\textit{y}\texttt{], 0, 1], }\textit{x}\texttt{]} \textit{stack}\ldots \\
\htablesize
$MERGE_4$  & \textit{none}                 & \texttt{if}      && \textit{y} \textit{x} \textit{stack}\ldots & $\rightarrow$ & \texttt{[4, }\textit{y}\texttt{, }\textit{x}\texttt{]} \textit{stack}\ldots \\
\htablesize
$SHARE$     & \texttt{int}                  & \textit{none}   && \textit{x} \textit{stack}\ldots & $\rightarrow$ & \textit{list-tail(x, arg)} \textit{x} \textit{stack}\ldots \\
\end{tabular}
\caption{The decoding instructions and their effect on the decoding stack.}
    \label{tab:decoding_instructions}
\end{table}

\endgroup

Table~\ref{tab:decoding_instructions} shows the decoding instructions
now supported.  On the right side is the effect of the decoding
instruction on the decoding stack state.  The top of stack is always a
rib and is a sequence of RVM instructions under construction.  The
$LINK$ instructions add an RVM instruction to the sequence (with no
change to the stack size). The RVM instruction added is either a
\texttt{call}, \texttt{set}, \texttt{get}, or \texttt{const} with an
\texttt{int} or \texttt{sym} operand.  The $MERGE$ instructions pop
one RVM code sequence from the stack and add either a \texttt{const}
or \texttt{if} RVM instruction to the now current topmost code
sequence (thus reducing the stack size by one). This allows
constructing a \texttt{const} RVM instruction referring to a constant
procedure whose arity is the argument of the $MERGE_3$ instruction
(this is often combined with a \texttt{set} RVM instruction to
implement top-level procedure definitions).  The $PUSH_0$ decoding
instruction pushes to the stack a new sequence containing a single
\texttt{jump} instruction with an \texttt{int} or \texttt{sym}
operand.  The $SHARE$ decoding instruction is the only other way to
start the construction of a code sequence. It extracts the tail of
the sequence currently under construction to start a new code sequence. It is used for each control
flow join point in the code graph. The argument is the number of
RVM instructions to skip. For example, when the code graph of
Figure~\ref{fig:abs} is converted to a RIBN a $SHARE$ decoding
instruction with an argument of 1 is used after the false branch of
the \texttt{if} has been constructed. Then the true branch is added
and a $MERGE_4$ decoding instruction is used to create the
\texttt{if}.

To determine what part of the code graph has sharing, a hash-consing
algorithm is used by the RSC compiler to construct the code
graph. Hash-consing can determine if two nodes, including all of their
children, are equal.  Equal code graph tails will automatically be
shared in the constructed code graph.  Other benefits also emerge from
hash-consing such as the ability to optimize certain forms of
duplication in the source code. For example, the two following
expressions will compile to the same code graph:

\vspace*{0.5ex}

\begin{center}
\begin{minipage}{2in}
\texttt{(if (< x y) (f x) (f y))} \\[0.3ex]
\texttt{(f (if (< x y) x y)))}
\end{minipage}
\end{center}

\subsection{Encoding of the Decoding Instructions}

\newcommand{\htablesize}[0]{\vspace{2pt}}

\begin{table}[]
    \begin{tabular}{llllll}
        \begin{tabular}[c]{@{}c@{}}
            RVM
        \end{tabular} & 
        \begin{tabular}[c]{@{}c@{}}
        \end{tabular} & 
        \begin{tabular}[c]{@{}c@{}}
        \end{tabular} & 
        \begin{tabular}[c]{@{}c@{}}
            Decoding
        \end{tabular} &
        \begin{tabular}[c]{@{}c@{}}
        \end{tabular} & 
        \begin{tabular}[c]{@{}c@{}}
        \end{tabular}  
        \\
        \begin{tabular}[c]{@{}c@{}}
            instruction
        \end{tabular} & 
        \begin{tabular}[c]{@{}c@{}}
            Range
        \end{tabular} & 
        \begin{tabular}[c]{@{}c@{}}
            Size
        \end{tabular} & 
        \begin{tabular}[c]{@{}c@{}}
            instruction
        \end{tabular} &
        \begin{tabular}[c]{@{}c@{}}
            Argument
        \end{tabular} & 
        \begin{tabular}[c]{@{}c@{}}
            Form
        \end{tabular}  
        \htablesize
        \\
        \htablesize
        \texttt{jump} & 0-19  & 20 &  $PUSH_0$     & \texttt{sym} & \textit{short} \\          
        \htablesize
        \texttt{jump} & 20-20  & 1 &  $PUSH_0$     & \texttt{int} & \textit{long} \\
        \htablesize
        \texttt{jump} & 21-22  & 2 &  $PUSH_0$     & \texttt{sym} & \textit{long} \\
        \htablesize
        \texttt{call} & 23-52  & 30 & $LINK_0$     & \texttt{sym} & \textit{short} \\
        \htablesize
        \texttt{call} & 53-53  & 1 &  $LINK_0$     & \texttt{int} & \textit{long} \\
        \htablesize
        \texttt{call} & 54-55  & 2 &  $LINK_0$     & \texttt{sym} & \textit{long} \\
        \htablesize
        \texttt{set}  & 56-56  & 1 &  $LINK_1$     & \texttt{int} & \textit{long} \\
        \htablesize
        \texttt{set}  & 57-59  & 2 &  $LINK_1$     & \texttt{sym} & \textit{long} \\
        \htablesize
        \texttt{get}  & 59-68  & 10 &  $LINK_2$    & \texttt{int} & \textit{short} \\
        \htablesize
        \texttt{get}  & 69-69  & 1  &  $LINK_2$    & \texttt{int} & \textit{long} \\
        \htablesize
        \texttt{get}  & 70-71  & 2  &  $LINK_2$    & \texttt{sym} & \textit{long} \\
        \htablesize
        \texttt{const} & 72-82  & 11  & $LINK_3$    & \texttt{int} & \textit{short} \\
        \htablesize
        \texttt{const} & 83-83  & 1  &  $LINK_3$    & \texttt{int} & \textit{long} \\
        \htablesize
        \texttt{const} & 84-85  & 2  &  $LINK_3$    & \texttt{sym} & \textit{long} \\
        \htablesize
        \texttt{const} & 86-89  & 4  &  $MERGE_3$   & \texttt{int} & \textit{short} \\
        \htablesize
        \texttt{const} & 90-90  & 1  &  $MERGE_3$   & \texttt{sym} & \textit{long} \\
        \htablesize
        \texttt{if} & 91-91  & 1  &  $MERGE_4$   &  &  \\
        \htablesize

    \end{tabular}
    \caption{Encoding of decoding instructions used in the previous Ribbit system.}
\label{tab:original_encoding}
\end{table}

The encoding of each decoding instruction contains its type, its argument's
type, and its argument's value. The argument's value can be encoded either with
a single RIBN code (\textit{short}) or across multiple ones (\textit{long}). For each
type of decoding instruction and argument, a range of codes is assigned. For the
\textit{short} encoding, the argument will be the difference between the code
and the lower boundary of the range. For the \textit{long} encoding, the
argument will utilize the VLQ encoding, with a starting value in the
accumulator equal to the difference between the code and the lower boundary of the
range. For instance, if a range is assigned to the codes \texttt{50}..\texttt{55} for
the \textit{long} encoding, then the two RIBN codes \texttt{53 4} will give the argument
3 $\times$ \emph{rb}/2 + 4.

In the previous Ribbit system, which utilized a fixed RIBN base of 92, the
ranges were as indicated in Table~\ref{tab:original_encoding}. These ranges
were determined by minimizing the RIBN size through a trial and error process on
the source code of the REPL. They are accessible to the decoder through a 
table indicating the size of the \textit{short} form of each decoding
instruction. For example, a RIBN code of 42 encodes a $LINK_0$ decoding
instruction with an argument of $19=42-23$. This generates a \texttt{call} RVM
instruction with a reference to the symbol at index 19 in the symbol table.

%

%

\subsection{Encoding Specialization}\label{sec:encoding_specialization}

Ribbit still uses the same encoding strategy but adapted to the RIBN base
of the RVM implementation and specialized to the code graph produced by
the compiler.

At compile time, the programmer can choose the encoding mechanism. The
\texttt{original} encoding represents the encoding in Table~\ref{tab:original_encoding}.
It has the advantages of being backward compatible and fast
to generate. The \texttt{optimal} encoding is a new approach
that searches for the best ranges for each decoding instruction for the
code graph generated. It
starts off with a range size of 1 for each \textit{long} encoding and a range size of 0
for the others. Then, greedily, it picks the best range to increase. The best
range is the one that has the best ratio between the number of bits
needed to encode the range and the number of bits saved by encoding this range
with a single RIBN code. Although it is not optimal in the theoretical sense
it gives good results in practice.

As the optimal encoding calculates, at compile time, a specialized encoding
for the code graph, the annotation system (see Section~\ref{sec:annotation})
has been extended to allow the replacement of information known by the
compiler. To do this, the \texttt{replace} annotation has been extended, as
explained in \ref{sec:ribbi_extension_replace}. Here is an example taken from
the JavaScript RVM:

\begin{verbatim}
// @@(feature encoding/optimal
while (1) {
    x = get_code();
    ...
    // @@(replace "[0,1,2]" (list->host encoding/optimal/start "[" "," "]")
    while((d=[0,1,2][++op]) <= n) n-=d 
    // )@@
    ...
// )@@
\end{verbatim}

In this code, an internal procedure of the compiler that is available in annotations is used to replace the source code \texttt{[0,1,2]}.
The procedure \texttt{list->host} takes a Scheme list, a prefix, a
separator and a suffix. It generates a string that concatenates the values
of the list, separated by the separator and surrounded by the prefix and
suffix. This is useful because this kind of syntax for a sequence of codes is almost
universal among programming languages. The feature \texttt{encoding/optimal/start} contains a
list of the start of the \textit{short} ranges for the optimal encoding. The order of the
decoding instructions is always the same for the optimal encoding, and thus known by the RVM. To
automatically choose the right decoder implementation in the RVM, the compiler
uses the feature system described in Section~\ref{sec:encoding}. The compiler
will activate the feature \texttt{encoding/optimal} when the optimal encoding
is used and the feature \texttt{encoding/original} when the original encoding
is used. This lets the RVM adjust its code to the encoding used.

\subsection{LZSS Compression} \label{sec:encoding_lzss}

LZSS~\cite{BELL86} is a general-purpose 
compression algorithm heavily inspired by LZ77~\cite{LZ77} that replaces recurring slices of text with back-pointers 
to previous occurrences.

The relative simplicity of LZSS makes it an interesting compression algorithm as the implementation of the decoder contributes to the total code size of the RVM.
Fancier algorithms like Zip or Bzip2 are much more complex and, unless the source program is very large, their implementation
will take more space than the space saved by the compression of the RIBN. A
LZSS decompressor is particularly compact; it has been implemented in less than 
50 lines of x86 assembly code and still offers effective compression of the RIBN.

The LZSS algorithm works on units of information that we will call
\emph{bytes} since in practice they correspond to 8 bits even though
in theory it could be different. The \emph{byte-base} (\emph{bb}) is the
number of codes in a \emph{byte}, i.e.~\emph{bb} = 256 in practice. The
decompressor takes a stream of bytes and outputs a sequence of RIBN
codes.

One of the key issues in applying the LZSS algorithm to compress the
RIBN is the encoding of back-pointers, which are composed of an
\emph{offset} and a \emph{size}. The \emph{offset} is the
backward distance in number of RIBN codes to the end of the section
that is repeated and \emph{size} is the length of the section.

To achieve good compression rates it is important to encode
back-pointers in as few bytes as possible. We chose to always use two
bytes.  When the decompressor encounters a byte whose value is lower
than the RIBN base (\emph{rb}), it represents a RIBN code that is output as
is. Otherwise the byte is in the \emph{compression-range} and it is
the first of the two bytes that encode a back-pointer \emph{BP}. The two
bytes are combined with the formula \emph{BP} = (\emph{byte1} -
\emph{rb}) $\times$ \emph{bb} + \emph{byte2}.  The
\emph{size} and \emph{offset} are then extracted using the
\emph{size base} (\emph{sb}):

\vspace*{0.5ex}

\begin{center}
\begin{minipage}{1.5in}
\emph{size} = \emph{BP} mod \emph{sb} + 3 \\
\emph{offset} = \emph{BP} div \emph{sb}
\end{minipage}
\end{center}

\vspace*{0.5ex}

\noindent
where div and mod are the integer division and modulo operators.
There is no gain in using back-pointers to encode repeated sequences
of \emph{size} 1 or 2, so the minimum \emph{size} is 3 and the maximum
\emph{size} is \emph{sb} + 2. The maximum \emph{offset} is
((\emph{bb} - \emph{rb} + 1) $\times$ \emph{bb} - 1) div \emph{sb}.

The value chosen for \emph{size base} determines the
balance between the range of \emph{sizes} and the range of
\emph{offsets} that can be encoded by a 2 byte back-pointer.
Given that the optimal \emph{size base} depends on the source
program it is RSC that determines the value. RSC will iterate
over a small reasonable range of values for \emph{size base} (7 to 13)
and picks the one that produces the best compression.
When using LZSS, a \emph{RIBN base} of 186 is used, as this gives
the best compression for the REPL.

%
%

%


\section{The R4RS Library}\label{sec:r4rs}

Ribbit needs to adhere to an official Scheme standard as to properly compare its 
implementation to other Scheme interpreters and compilers. In order to stay tuned
with Ribbit's minimalism, the chosen standard needs to offer a good balance 
between expressiveness and a relatively small number of essential features. The 
R4RS standard is a good fit for Ribbit as it possesses such qualities. 

The two standards that were considered were R4RS and R5RS, as the 
others are simply too big (R6RS and R7RS) or too outdated (R3RS and below).
After analyzing the two, it becomes clear that the R4RS standard is a more 
sound choice for Ribbit compared to the R5RS standard for three primary reasons. 
First, R4RS is almost a subset of R5RS in terms of essential procedures and 
syntax (the only exception is the \texttt{load} procedure essential in R4RS but optional in
R5RS). Second, the R4RS standard defines $164$ essential procedures and 
$18$ essential syntaxes while the R5RS standard defines $207$ essential procedures 
and $26$ essential syntaxes. When size matters, this difference becomes significant. 
Third, hygienic macros, which are a key feature of R5RS, require a considerable amount of 
space to implement. Incorporating support for this into the \texttt{eval} procedure of the library would increase its
complexity and size beyond necessity.

\subsection{Design Choices and Tradeoffs}

Ribbit's instruction graph is composed of ribs that can either contain signed
integers or a reference to a rib. This has a number of consequences on the
implementation of the R4RS standard, especially when it comes to the
definitions of the various data types, as each type needs to adhere to this
structure.

One of the first constraints is that Ribbit only supports exact signed integers
inside the Scheme source code. The biggest challenge to support floating point
numbers is the variety of RVMs that Ribbit runs on. Ribbit could not rely on
the host's implementation as the behavior would not be consistent across RVMs. 
For example, the floating point arithmetic of an RVM written in POSIX Shell will behave differently 
than its counterpart in Python, making this behavior unreliable, which defeats the 
whole point of supporting multiple host languages. This means that Ribbit would need 
to implement its own floating point arithmetic, which increases the overall footprint 
size of the RVMs. To limit the efficiency loss, Ribbit could use the host's floating 
point in some cases and its own in others. However, this requires a lot of work 
for a feature that is optional in the R4RS standard.

In the previous Ribbit system, individual characters were simply represented as 
their integer code. However, the R4RS standard requires the adherence to the
\emph{Disjointedness of types principle}, meaning the type for characters must now be 
distinct from other types. For this reason, Ribbit now supports, behind a compiler 
feature named \verb|chars|, the representation of characters using a rib. The character
rib has the character code (an integer) in the first field and the type code 6
in the third field. The second field is an implementation dependent value and is 
currently unused by Ribbit's implementation of R4RS. This change makes the
creation of character constants require significantly more instructions than previously. 
This explains a common optimization done in procedures dealing with characters, 
like \verb|char>?| and \verb|char-whitespace?|, where characters are 
unboxed and the logic is done with their integer representation. To alleviate the
problem of size and avoid frequent character allocation, the use of a table of previously 
created characters was considered. This would also make characters comparison for 
equality quick using a \verb|eq?| test. However, in the interest of a simpler implementation, 
characters are always allocated when needed, for example by the \verb|integer->char|, 
\verb|read-char|, and \verb|string-ref| procedures. This also means that the procedure 
\verb|eqv?| needs to handle this case in the Scheme code, as the \verb|##eqv?| 
primitive only guarantees arithmetic equality, as with the \verb|=| procedure, for 
numbers and a reference equality, as with the \verb|eq?| procedure, for ribs. 

The change to the representation of characters doesn't impact the internal representation 
of strings however, as they were already differentiable from the other types. 
Therefore, strings are still represented as a rib containing a list of integer 
character codes in the first field, the length of the string in the second field, 
and the string type tag in the third field. While this representation must be 
accounted for in the implementation of some R4RS procedures, like 
\verb|string-ref| and \verb|string->list|, it also enables some optimizations
elsewhere, like in string comparisons with \verb|string<?|, \verb|string>?|, etc.

%
%
%
%

\subsection{A Portable I/O System}

The way of interfacing with I/O varies greatly between languages. To solve this problem,
Ribbit reduces the responsibilities of each RVM to a minimum by implementing most of 
the logic in Scheme code directly. This allows Ribbit to present a unified 
API that can adapt to all RVMs while complying with the behaviors expected by R4RS.

Ribbit separates \verb|input-port| and \verb|output-port| into two distinct 
data types, as required by R4RS.

An \verb|input-port| is a rib with this layout: \texttt{[}\emph{fd}\texttt{,} \emph{peeked-char/open?}\texttt{,} \texttt{8]}.
The \emph{fd} field is reserved by the RVM for implementation dependent file
descriptors. Each host language has its own way of communicating with the file
system, and this object bridges the gap between the host language and the RVM.
For example, the x86 assembly implementation uses an integer representing the 
Linux file descriptor, while the Python implementation keeps a reference to
a Python \verb|File| object. 

The \emph{peeked-char/open?} field has two purposes as to avoid using an 
extra rib. Its first role is for caching the last read character after 
a peek. It is used in the implementation of \verb|peek-char| and \verb|read-char|. 
It is needed because the RVM's file interface does not necessarily conform to
the R4RS standard. This field is the empty list by default, meaning there is no 
character peeked. The second role of the field is to indicate if the port is open or closed. 
To do so, the field is set to \verb|#f| when the port is closed and checking the state 
of a port becomes a simple \verb|not| test. It is necessary to provide this information
as R4RS mandates that a port may be closed any number of times without causing an error,
a guarantee not shared by all languages. For example, NodeJS will throw an exception while 
attempting to close an already closed port.

An \verb|output-port| is a rib with this layout: \texttt{[}\emph{fd}\texttt{,} \emph{open?}\texttt{,} \texttt{9]}.
The \emph{fd} field and the \emph{open?} field have the same meaning as those
fields in \verb|input-port|.

To support I/O the following primitive procedures must be defined by an RVM:
\begin{itemize}
    \item[] \texttt{(\#\#stdin-fd)}
    \item[] \texttt{(\#\#stdout-fd)}
    \item[] \texttt{(\#\#get-fd-input-file} \emph{filename}\texttt{)}
    \item[] \texttt{(\#\#get-fd-output-file} \emph{filename}\texttt{)}
    \item[] \texttt{(\#\#read-char-fd} \emph{fd}\texttt{)}
    \item[] \texttt{(\#\#write-char-fd} \emph{ch-code} \emph{fd}\texttt{)}
    \item[] \texttt{(\#\#close-input-fd} \emph{fd}\texttt{)}
    \item[] \texttt{(\#\#close-output-fd} \emph{fd}\texttt{)}
\end{itemize}

\verb|##stdin-fd| and \verb|##stdout-fd| return the \emph{fd} host-dependent value used in the standard input
and output port. \verb|##get-fd-input-file| and \verb|##get-fd-output-file| take a filename
and return the \emph{fd} host-dependent value used in the input and output port for that file.
\verb|##read-char-fd| takes a \emph{fd} and reads a character code (as an integer) 
from the file, while \verb|##write-char-fd| takes a character code (as an integer) 
and a \emph{fd} and writes the character to the file. Finally, \verb|##close-input-fd| 
and \verb|##close-output-fd| take a \emph{fd} and close the corresponding port. 
All those procedures either take or return the implementation dependent object 
\emph{fd} that is used to retrieve or write data to the file system using the 
host language.

Scheme procedures are defined on top of these primitives, and they take care of
caching the peeked character, checking if the port is open, closing the port when 
necessary, returning the \verb|end-of-file| object, and converting the character code (integer)
read into a Scheme character.

\subsection{Strategies Used for Making a Compact R4RS Library}

The Ribbit implementation of the R4RS library is optimized for size at the cost of
execution speed. This approach can be seen in the implementation of 
\verb|+|, \verb|*|, and \verb|-|, which are all defined using the \verb|fold|
procedure, as a call to \verb|fold| takes less space than an explicit loop to 
compute the sum or product. The \verb|fold| procedure, even if absent from the R4RS standard and, therefore,
not required, is used often enough to make up for the space its implementation 
needs. The pattern of using higher order procedures to generalize a behavior is 
used frequently in the R4RS implementation, as it often avoids code repetition.

Another trick is to define procedures in terms of other procedures. For example,
\verb|(char>? c1 c2)| is defined as \verb|(char<? c2 c1)| instead of the 
faster, but larger \verb|(> (char->integer c1) (char->integer c2))|.

Another common technique is to define a few internal procedures of a general nature and to call them in many different places. This makes the AOT compiler 
rank those procedures at low indexes in the symbol-table, reducing the cost of 
accessing them to, in the best case, as little as one byte.

\subsection{Expander Macros}

The Ribbit AOT compiler supports
the \verb|define-expander| special form for defining \emph{expander macros} that are
responsible for handling their own recursive expansion \cite{GENERALMACRO}. This is used in
the R4RS implementation to optimize certain common patterns.
For example, a call to the procedure \verb|+| with two arguments will be expanded to 
a call of the \verb|##+| primitive, which avoids a call to \verb|fold|. Expander macros
are used extensively by the implementation of R4RS to improve execution speed without compromising space.

\subsection{Testing the R4RS Compliance of the Compiler and REPL}

Compliance to R4RS was verified using a series of tests inspired by, or taken from, 
the R4RS test file of Chicken \cite{CHICKTEST, CHICKEN}, which includes many
examples from the R4RS document. The
use of multiple test files provides modular compliance testing.
Each test file starts with the code tested followed by
comments containing the expected output.
These were run on the JavaScript, Python, and x86 assembly RVM.

To test the Ribbit compiler, the makefile iterates over the test files and for each one:
\begin{enumerate}
    \item Compiles the test file to the target host using the Ribbit compiler;
    \item Runs the generated RVM using the host interpreter or compiler;
    \item Compares the values written to the standard output to
          the expected output.
\end{enumerate}

To test the Ribbit R4RS REPL, the makefile:
\begin{enumerate}
    \item Compiles the REPL to the target host using the Ribbit compiler;
    \item Runs the generated RVM using the host interpreter or compiler;
    \item For each test file, evaluates \verb|(load "path/to/the/test.scm")| and compares 
      the values written to the standard output to the expected output.
\end{enumerate}

\section{X86 Assembly Host}\label{sec:x86-host}

Choosing the right host is critical for implementing the R4RS standard within a
7 KB limit.  While the library is designed to work on any host, achieving the
goal of a 7 KB footprint requires optimization efforts focused specifically on
the host.

High-level languages like Python or JavaScript could serve as intriguing host
options. The footprint of these hosts is determined by the size of the source
code for the resulting RVM. Since users are likely to have the host language
pre-installed, only the source code is required to run the RVM. The challenge
here lies in balancing the verbosity of the source code against the availability
of built-in language features. 

The C language is an appealing host option due to its compactness and ease of
optimization to meet the 7 KB size constraint. However, executables generated
through \texttt{gcc} often include unnecessary boilerplate code, like the main
function startup code. While the generated code can be trimmed-down with C compiler
options, using C as a host still limits our low-level control~\cite{tuto}.

A stripped down ELF file containing x86 code is our host of choice. It is very
compact and optimizable to meet our objectives. However, everything needs to be implemented from scratch,
including a GC, the I/O primitives (using Linux syscalls) and the specialized encoding.  Fortunately,
the RVM is sufficiently simple to allow for the implementation of this kind of low-level host within a
reasonable time frame.

\section{Evaluation}\label{sec:evaluation}

We are interested in measuring the footprint of our R4RS implementation as a
standalone executable as well as the execution speed. The footprint of the REPL, including the generated RVM
for the x86 assembly host, is shown in Table~\ref{tab:footprint}. Each column
represents a different compilation setting. The first column is the baseline, which
is the size of the generated code without any optimizations. The available
compilation settings are:

\begin{description}
     
    \item[Baseline.] The \texttt{original} encoding is used.

    \item[Prim-no-arity.] The procedure call argument count is normally pushed to the
        stack, costing one byte per encoded call. If rest parameters are not
        used with primitives, this can be skipped for primitives. The space saving is
        appreciable as all calls to primitives are encoded with one fewer byte and the argument count check in the RVM can be removed if rest parameters are not used in the source program.

    \item[Optimal (92).] The \texttt{optimal} encoding with 92 codes per byte is used, as described in
        Section~\ref{sec:encoding_specialization}.

    \item[Optimal (256).] The \texttt{optimal} encoding with 256 codes per byte is used,
        as described in Section~\ref{sec:encoding_specialization}.

    \item[LZSS.] LZSS compression is applied to the generated code
        before writing it to the RVM as described in Section~\ref{sec:encoding_lzss}. Note
        that LZSS compression works only with optimal (256) encoding.
        
\end{description}

The most compact executable for the REPL is obtained, unsurprisingly, with the combination of Prim-no-arity, Optimal (256), and LZSS: a footprint of 6.5 KB. Not using Prim-no-arity has a minor 2\% impact on footprint when using LZSS.

\begin{table}[h!]
\small


\begin{tabular}{llllllllll}
  &
  Baseline &
  Prim-no-arity &
  Optimal (92) &
  Optimal (256) &
  \begin{tabular}[c]{@{}l@{}}Prim-no-arity\\ Optimal (92)\end{tabular} &
  \begin{tabular}[c]{@{}l@{}}Prim-no-arity\\ Optimal (256)\end{tabular} \\

  No LZSS &
  14   KB &
  13   KB &
  9.8  KB &
  9.2  KB &
  9.0  KB & 
  8.4  KB \\
  
  LZSS &
  -    &
  -    &
  -    &
  6.6  KB &
  -    &
  6.5  KB \\
  
\end{tabular}
\caption{Footprint of the REPL compiled to the x86 assembly host with different compilation settings.}
\label{tab:footprint}
\end{table}

\newcommand{\secs}[1]{\footnotesize{\textbf{#1}~s}}
\newcommand{\kb}[1]{\footnotesize{#1~KB}}
\newcommand{\factor}[1]{\footnotesize{\textbf{#1$\times$}}}

To test the footprint and execution speed for specific source programs, the x86 assembly and the JavaScript host have been benchmarked, as well as the Gambit Scheme Interpreter~\cite{gambit}. Tests are taken from the Gambit benchmarking suite~\cite{gambit}. The test machine is a 4.5 GHz Intel i7-9750H with 16 GB of RAM running Linux. The NodeJS version is v10.24.1. The Gambit version is v4.9.5. In the Tables~\ref{tab:x86} and \ref{tab:nodejs}, different settings were used to compile/execute the benchmark: 

\begin{description}
	\item[gsi.] Gambit Scheme Interpreter used as a reference for execution speed comparison.
	\item[pna.] Refers to the Prim-no-arity compilation option. This is the same as in Table~\ref{tab:footprint}. Note that we use the label \textbf{pa} to indicate not using that option (i.e.~the primitives do check arity).
	\item[tc.] The R4RS library comes in two forms: with and without type checking. The \texttt{tc} option means that the type checking version is used. Note that type checking is not required for conformance with R4RS.
	\item[x86 REPL.] Execution using the REPL compiled for the x86 assembly host. It was compiled with Prim-no-arity, Optimal (256), LZSS, and the non type checking R4RS library (footprint of 6.5 KB).
\end{description}

\begin{table}[h!]
\begin{tabular}{|l|r|rr|rr|rr|rr|r|}
\cline{2-11}
\multicolumn{1}{l|}{} 
& \multicolumn{1}{c|}{\textbf{gsi}} 
& \multicolumn{2}{c|}{} 
& \multicolumn{2}{c|}{} 
& \multicolumn{2}{c|}{\textbf{tc}} 
& \multicolumn{2}{c|}{\textbf{tc}} 
& \multicolumn{1}{c|}{\textbf{x86}} 
\\
\multicolumn{1}{l|}{} 
& \multicolumn{1}{c|}{(secs)} 
& \multicolumn{2}{c|}{\textbf{pna}} 
& \multicolumn{2}{c|}{\textbf{pa}} 
& \multicolumn{2}{c|}{\textbf{pna}} 
& \multicolumn{2}{c|}{\textbf{pa}} 
& \multicolumn{1}{c|}{\textbf{REPL}} 


 \\ 
\hline
\multicolumn{1}{|l|}{ctak} & \multicolumn{1}{r|}{\secs{0.2}} & \multicolumn{1}{r}{\factor{0.9}} & \multicolumn{1}{r|}{\kb{2.1}} & \multicolumn{1}{r}{\factor{1.0}} & \multicolumn{1}{r|}{\kb{2.2}} & \multicolumn{1}{r}{\factor{1.5}} & \multicolumn{1}{r|}{\kb{8.8}} & \multicolumn{1}{r}{\factor{1.8}} & \multicolumn{1}{r|}{\kb{9.3}} & \multicolumn{1}{r|}{\factor{2.8}}  \\ \hline
\multicolumn{1}{|l|}{fib} & \multicolumn{1}{r|}{\secs{26.2}} & \multicolumn{1}{r}{\factor{0.3}} & \multicolumn{1}{r|}{\kb{2.0}} & \multicolumn{1}{r}{\factor{0.4}} & \multicolumn{1}{r|}{\kb{2.0}} & \multicolumn{1}{r}{\factor{1.9}} & \multicolumn{1}{r|}{\kb{8.6}} & \multicolumn{1}{r}{\factor{2.2}} & \multicolumn{1}{r|}{\kb{9.1}} & \multicolumn{1}{r|}{\factor{4.9}}  \\ \hline
\multicolumn{1}{|l|}{ack} & \multicolumn{1}{r|}{\secs{2.2}} & \multicolumn{1}{r}{\factor{0.4}} & \multicolumn{1}{r|}{\kb{2.0}} & \multicolumn{1}{r}{\factor{0.4}} & \multicolumn{1}{r|}{\kb{2.0}} & \multicolumn{1}{r}{\factor{1.6}} & \multicolumn{1}{r|}{\kb{8.6}} & \multicolumn{1}{r}{\factor{1.9}} & \multicolumn{1}{r|}{\kb{9.1}} & \multicolumn{1}{r|}{\factor{5.7}}  \\ \hline
\multicolumn{1}{|l|}{tak} & \multicolumn{1}{r|}{\secs{2.3}} & \multicolumn{1}{r}{\factor{0.6}} & \multicolumn{1}{r|}{\kb{2.0}} & \multicolumn{1}{r}{\factor{0.6}} & \multicolumn{1}{r|}{\kb{2.0}} & \multicolumn{1}{r}{\factor{1.5}} & \multicolumn{1}{r|}{\kb{8.6}} & \multicolumn{1}{r}{\factor{1.7}} & \multicolumn{1}{r|}{\kb{9.1}} & \multicolumn{1}{r|}{\factor{3.5}}  \\ \hline
\multicolumn{1}{|l|}{takl} & \multicolumn{1}{r|}{\secs{2.5}} & \multicolumn{1}{r}{\factor{0.9}} & \multicolumn{1}{r|}{\kb{2.2}} & \multicolumn{1}{r}{\factor{1.0}} & \multicolumn{1}{r|}{\kb{2.2}} & \multicolumn{1}{r}{\factor{0.8}} & \multicolumn{1}{r|}{\kb{8.7}} & \multicolumn{1}{r}{\factor{1.0}} & \multicolumn{1}{r|}{\kb{9.2}} & \multicolumn{1}{r|}{\factor{1.0}}  \\ \hline
\multicolumn{1}{|l|}{primes} & \multicolumn{1}{r|}{\secs{1.4}} & \multicolumn{1}{r}{\factor{0.8}} & \multicolumn{1}{r|}{\kb{2.3}} & \multicolumn{1}{r}{\factor{1.0}} & \multicolumn{1}{r|}{\kb{2.3}} & \multicolumn{1}{r}{\factor{1.5}} & \multicolumn{1}{r|}{\kb{8.8}} & \multicolumn{1}{r}{\factor{1.8}} & \multicolumn{1}{r|}{\kb{9.3}} & \multicolumn{1}{r|}{\factor{2.6}}  \\ \hline
\multicolumn{1}{|l|}{deriv} & \multicolumn{1}{r|}{\secs{0.7}} & \multicolumn{1}{r}{\factor{6.8}} & \multicolumn{1}{r|}{\kb{2.7}} & \multicolumn{1}{r}{\factor{8.2}} & \multicolumn{1}{r|}{\kb{2.7}} & \multicolumn{1}{r}{\factor{26.3}} & \multicolumn{1}{r|}{\kb{9.2}} & \multicolumn{1}{r}{\factor{32.6}} & \multicolumn{1}{r|}{\kb{9.8}} & \multicolumn{1}{r|}{\factor{7.3}}  \\ \hline
\multicolumn{1}{|l|}{mazefun} & \multicolumn{1}{r|}{\secs{1.4}} & \multicolumn{1}{r}{\factor{0.7}} & \multicolumn{1}{r|}{\kb{4.0}} & \multicolumn{1}{r}{\factor{0.8}} & \multicolumn{1}{r|}{\kb{4.1}} & \multicolumn{1}{r}{\factor{1.7}} & \multicolumn{1}{r|}{\kb{9.9}} & \multicolumn{1}{r}{\factor{2.0}} & \multicolumn{1}{r|}{\kb{11}} & \multicolumn{1}{r|}{N/A}  \\ \hline
\multicolumn{1}{|l|}{nqueens} & \multicolumn{1}{r|}{\secs{2.0}} & \multicolumn{1}{r}{\factor{0.8}} & \multicolumn{1}{r|}{\kb{3.4}} & \multicolumn{1}{r}{\factor{0.9}} & \multicolumn{1}{r|}{\kb{3.5}} & \multicolumn{1}{r}{\factor{1.7}} & \multicolumn{1}{r|}{\kb{8.8}} & \multicolumn{1}{r}{\factor{2.0}} & \multicolumn{1}{r|}{\kb{9.2}} & \multicolumn{1}{r|}{N/A}  \\ \hline
\multicolumn{1}{|l|}{sum} & \multicolumn{1}{r|}{\secs{19.4}} & \multicolumn{1}{r}{\factor{0.3}} & \multicolumn{1}{r|}{\kb{2.0}} & \multicolumn{1}{r}{\factor{0.4}} & \multicolumn{1}{r|}{\kb{2.0}} & \multicolumn{1}{r}{\factor{2.1}} & \multicolumn{1}{r|}{\kb{8.6}} & \multicolumn{1}{r}{\factor{2.5}} & \multicolumn{1}{r|}{\kb{9.1}} & \multicolumn{1}{r|}{N/A}  \\ \hline

\end{tabular}
\caption{Execution time when using the Gambit Scheme Interpreter and for Ribbit the relative execution time and footprint of the x86 assembly host on different benchmarks.}
\label{tab:x86}
\end{table}

\begin{table}[h!]
\begin{tabular}{|l|r|rr|rr|rr|rr|}
\cline{2-10}
\multicolumn{1}{l|}{} 
& \multicolumn{1}{c|}{\textbf{gsi}} 
& \multicolumn{2}{c|}{} 
& \multicolumn{2}{c|}{} 
& \multicolumn{2}{c|}{\textbf{tc}} 
& \multicolumn{2}{c|}{\textbf{tc}} 
\\
\multicolumn{1}{l|}{} 
& \multicolumn{1}{c|}{(secs)} 
& \multicolumn{2}{c|}{\textbf{pna}} 
& \multicolumn{2}{c|}{\textbf{pa}} 
& \multicolumn{2}{c|}{\textbf{pna}} 
& \multicolumn{2}{c|}{\textbf{pa}} 
 \\ 
\hline

\multicolumn{1}{|l|}{ctak} & \multicolumn{1}{r|}{\secs{0.2}} & \multicolumn{1}{r}{\factor{10.6}} & \multicolumn{1}{r|}{\kb{2.7}} & \multicolumn{1}{r}{\factor{13.2}} & \multicolumn{1}{r|}{\kb{2.7}} & \multicolumn{1}{r}{\factor{18.3}} & \multicolumn{1}{r|}{\kb{11}} & \multicolumn{1}{r}{\factor{22.0}} & \multicolumn{1}{r|}{\kb{11}} \\ \hline
\multicolumn{1}{|l|}{fib} & \multicolumn{1}{r|}{\secs{26.2}} & \multicolumn{1}{r}{\factor{4.0}} & \multicolumn{1}{r|}{\kb{2.4}} & \multicolumn{1}{r}{\factor{5.1}} & \multicolumn{1}{r|}{\kb{2.5}} & \multicolumn{1}{r}{FAIL} & \multicolumn{1}{r|}{\kb{11}} & \multicolumn{1}{r}{FAIL} & \multicolumn{1}{r|}{\kb{11}} \\ \hline
\multicolumn{1}{|l|}{ack} & \multicolumn{1}{r|}{\secs{2.2}} & \multicolumn{1}{r}{\factor{4.8}} & \multicolumn{1}{r|}{\kb{2.4}} & \multicolumn{1}{r}{\factor{5.3}} & \multicolumn{1}{r|}{\kb{2.5}} & \multicolumn{1}{r}{\factor{21.5}} & \multicolumn{1}{r|}{\kb{11}} & \multicolumn{1}{r}{\factor{22.9}} & \multicolumn{1}{r|}{\kb{11}} \\ \hline
\multicolumn{1}{|l|}{tak} & \multicolumn{1}{r|}{\secs{2.3}} & \multicolumn{1}{r}{\factor{7.0}} & \multicolumn{1}{r|}{\kb{2.4}} & \multicolumn{1}{r}{\factor{8.3}} & \multicolumn{1}{r|}{\kb{2.5}} & \multicolumn{1}{r}{\factor{20.4}} & \multicolumn{1}{r|}{\kb{11}} & \multicolumn{1}{r}{\factor{23.1}} & \multicolumn{1}{r|}{\kb{11}} \\ \hline
\multicolumn{1}{|l|}{takl} & \multicolumn{1}{r|}{\secs{2.5}} & \multicolumn{1}{r}{\factor{11.5}} & \multicolumn{1}{r|}{\kb{2.7}} & \multicolumn{1}{r}{\factor{13.5}} & \multicolumn{1}{r|}{\kb{2.7}} & \multicolumn{1}{r}{\factor{12.5}} & \multicolumn{1}{r|}{\kb{11}} & \multicolumn{1}{r}{\factor{13.7}} & \multicolumn{1}{r|}{\kb{11}} \\ \hline
\multicolumn{1}{|l|}{primes} & \multicolumn{1}{r|}{\secs{1.4}} & \multicolumn{1}{r}{\factor{10.6}} & \multicolumn{1}{r|}{\kb{2.8}} & \multicolumn{1}{r}{\factor{12.3}} & \multicolumn{1}{r|}{\kb{2.9}} & \multicolumn{1}{r}{\factor{19.7}} & \multicolumn{1}{r|}{\kb{11}} & \multicolumn{1}{r}{\factor{25.0}} & \multicolumn{1}{r|}{\kb{11}} \\ \hline
\multicolumn{1}{|l|}{deriv} & \multicolumn{1}{r|}{\secs{0.7}} & \multicolumn{1}{r}{\factor{94.2}} & \multicolumn{1}{r|}{\kb{3.3}} & \multicolumn{1}{r}{\factor{108.0}} & \multicolumn{1}{r|}{\kb{3.4}} & \multicolumn{1}{r}{FAIL} & \multicolumn{1}{r|}{\kb{11}} & \multicolumn{1}{r}{FAIL} & \multicolumn{1}{r|}{\kb{12}} \\ \hline
\multicolumn{1}{|l|}{mazefun} & \multicolumn{1}{r|}{\secs{1.4}} & \multicolumn{1}{r}{\factor{8.5}} & \multicolumn{1}{r|}{\kb{5.0}} & \multicolumn{1}{r}{\factor{10.5}} & \multicolumn{1}{r|}{\kb{5.2}} & \multicolumn{1}{r}{\factor{23.8}} & \multicolumn{1}{r|}{\kb{12}} & \multicolumn{1}{r}{\factor{28.0}} & \multicolumn{1}{r|}{\kb{13}} \\ \hline
\multicolumn{1}{|l|}{nqueens} & \multicolumn{1}{r|}{\secs{2.0}} & \multicolumn{1}{r}{\factor{10.2}} & \multicolumn{1}{r|}{\kb{4.2}} & \multicolumn{1}{r}{\factor{11.8}} & \multicolumn{1}{r|}{\kb{4.4}} & \multicolumn{1}{r}{\factor{24.7}} & \multicolumn{1}{r|}{\kb{11}} & \multicolumn{1}{r}{\factor{27.6}} & \multicolumn{1}{r|}{\kb{11}} \\ \hline
\multicolumn{1}{|l|}{sum} & \multicolumn{1}{r|}{\secs{19.4}} & \multicolumn{1}{r}{\factor{4.2}} & \multicolumn{1}{r|}{\kb{2.4}} & \multicolumn{1}{r}{\factor{5.3}} & \multicolumn{1}{r|}{\kb{2.5}} & \multicolumn{1}{r}{FAIL} & \multicolumn{1}{r|}{\kb{11}} & \multicolumn{1}{r}{FAIL} & \multicolumn{1}{r|}{\kb{11}} \\ \hline
\end{tabular}
\caption{Execution time when using the Gambit Scheme Interpreter and for Ribbit the relative execution time and footprint of the NodeJS host on different benchmarks.}
\label{tab:nodejs}
\end{table}

Note that a few entries are missing from the tables.
Some of the benchmarks use named-\texttt{let} and internal \texttt{define} which are not supported by the REPL because they are not required by R4RS.
This is indicated with N/A in Table~\ref{tab:x86}.
In Table~\ref{tab:nodejs}, a few tests with the type checked version timed-out. The limit was 5 minutes for the compilation and the execution of the benchmark.

The results for the x86 assembly host demonstrate good space and execution speed
characteristics when the programs are compiled with the AOT compiler.
These relatively small benchmark programs (15-200 LOC) compile to executables in the 2-4 KB range when the non type checking R4RS library is used. It demonstrates the effectiveness of the AOT compiler to remove unused parts of the R4RS library and RVM source code. In terms of execution speed.
When the type checking R4RS library is used the footprint grows considerably to the 9-11 KB range. This is due to the frequent use of \verb|set!| to redefine the predefined procedures with type checking variants which interferes with the effectiveness of the dead code elimination.

Execution speed compares well with the Gambit Scheme Interpreter. All programs except \verb|deriv| are faster when compiled with the Ribbit AOT compiler and non type checking library is used.  The AOT compiler does not optimize \verb|deriv| well due to the presence of higher-order procedures and rest parameters.
The x86 REPL fares reasonably well in terms of execution speed with a factor of no more than 7.3$\times$ slower than the Gambit Scheme Interpreter.

The results for the JavaScript host show that the footprint is consistently about 0.5 KB larger than with the x86 assembly host. On the other hand, the execution time is about one order of magnitude larger, which can be explained by the use of a higher level host language without low level control.


\section{Related Work}\label{sec:related}

Bit~\cite{BIT} is a compact Scheme implementation based on an AOT compiler. It supports \texttt{call/cc}, and most
constructs from R4RS. However, it doesn't support the full R4RS standard,
excluding all port based textual I/O procedures. It claims to fit this implementation within 22 KB.
In contrast, Ribbit is a factor of 3$\times$ smaller while providing a REPL and a fully compliant
R4RS library.

Picobit~\cite{PICOBIT} is a Scheme implementation also based on an AOT compiler
that targets embedded systems. It supports a broad subset of R5RS and includes
a macro system, however,
important features are missing: file I/O, \texttt{eval}, and string-to-symbol conversion. They claim to fit the
VM without the standard library and without bignums in 11.6 KB
on PIC18 microcontrollers which are 8-bit microprocessors.
Moreover the heap size is constrained by the use of a 16 bit address space.
Ribbit has a substantially smaller VM (\~{ }2 KB for the x86 RVM which has a 32 bit address space) and the AOT compiler supports macros.
Additionally, Ribbit offers a full REPL compliant with
R4RS, including support for string-to-symbol conversion, \texttt{eval}, and I/O
procedures, all within 6.5 KB which is considerably smaller than the bare Picobit VM.

Bigloo~\cite{BIGLOO} is a Scheme implementation with a macro system that is very similar to the one used by Ribbit, as they both 
define a \verb|define-expander| special form with a similar semantics. For example, in
Bigloo, as well as in Ribbit, the form \verb|define-macro| is expanded into a \verb|define-expander|.
A notable difference is that Ribbit only supports the use of expanders in the Ribbit AOT compiler, while 
Bigloo supports this in its compiler and interpreter.

SectorLISP~\cite{sectorlisp} serves as a unique Lisp implementation that runs directly as an operating system with a GC and impressively fits within the constraints of a 512-byte boot sector. While this compactness is noteworthy, it comes with limitations in terms of features when compared to Ribbit. Specifically, SectorLISP does not include a built-in \texttt{eval} procedure - though users can manually enter one through the REPL - and its Lisp version falls short of the comprehensive feature set found in the R4RS standard.

\section{Conclusion}\label{sec:conclusion}

In this paper, we have described how the Ribbit system has been improved with
a new, more efficient, and more flexible way of encoding Scheme programs. We also have
described how a R4RS compliant REPL was implemented for Ribbit in a footprint 
of only 6.5 KB. The REPL is capable of running on a wide variety of host languages with no extra
dependencies. Our approach to minimize the generated encoded program by the Ribbit AOT compiler
is to use a multistep process. First, we do a liveness analysis on the Scheme code to remove any 
unused procedures and variables. The liveness analysis is also used to determine 
which parts of the generated Ribbit VM need to be removed, using our \verb|feature| system.
Then, the compiler finds the optimal way of encoding the program. After being encoded, 
the LZSS algorithm is used to compress it to reduce its size even more. Finally, the 
compiler injects the encoded program into the Ribbit VM.

It would be interesting to see how the system needs to be extended to support Scheme standards beyond R4RS. While we can only speculate on the compactness, we believe it is likely that a complete R7RS compliant REPL can be implemented in a 15-30 KB footprint.



\bibliographystyle{ACM-Reference-Format}
\bibliography{bio}
\end{document}